\documentclass[reprint,longbibliography,superscriptaddress]{revtex4-1}
\usepackage{amsmath,graphicx,hyperref,upgreek,stfloats,bm,comment}
\begin{document}
\title{
All-optical sub-Kelvin sympathetic cooling of a levitated microsphere in vacuum
}

\author{Yoshihiko Arita} \email{ya10@st-andrews.ac.uk}
\affiliation{SUPA, School of Physics \& Astronomy, University of St Andrews, North Haugh, St Andrews, KY16 9SS, United Kingdom}

\author{Graham D. Bruce}
\affiliation{SUPA, School of Physics \& Astronomy, University of St Andrews, North Haugh, St Andrews, KY16 9SS, United Kingdom}

\author{Ewan M.~Wright}
\affiliation{Wyant College of Optical Sciences, The University of Arizona, 1630 East University Boulevard, Tucson, Arizona 85721, USA}
\affiliation{SUPA, School of Physics \& Astronomy, University of St Andrews, North Haugh, St Andrews, KY16 9SS, United Kingdom}

\author{Stephen H. Simpson}
\affiliation{Institute of Scientific Instruments of the Czech Academy of Science, v.v.i., Kr\'alovopolsk\'a 147, 612 64 Brno, Czech Republic}

\author{Pavel Zem\'{a}nek}
\affiliation{Institute of Scientific Instruments of the Czech Academy of Science, v.v.i., Kr\'alovopolsk\'a 147, 612 64 Brno, Czech Republic}

\author{Kishan Dholakia} \email{kd1@st-andrews.ac.uk}
\affiliation{SUPA, School of Physics \& Astronomy, University of St Andrews, North Haugh, St Andrews, KY16 9SS, United Kingdom}
\affiliation{Wyant College of Optical Sciences, The University of Arizona, 1630 East University Boulevard, Tucson, Arizona 85721, USA}
\affiliation{Department of Physics, College of Science, Yonsei University, Seoul 03722, South Korea}
\affiliation{School of Biological Sciences, The University of Adelaide, Adelaide, South Australia, Australia}

\begin{abstract}
\noindent We demonstrate all-optical sympathetic cooling of a laser-trapped microsphere to sub-Kelvin temperatures, mediate by optical binding to a feedback-cooled adjacent particle. Our study opens prospects for multi-particle quantum entanglement and sensing in levitated optomechanics.

 \bigskip

\emph{This article was published by Optica Publishing Group under the terms of the \href{https://creativecommons.org/licenses/by/4.0/}{Creative
Commons Attribution 4.0 License}. Further distribution of this work must
maintain attribution to the author(s) and the published article’s title, journal
citation, and DOI \href{https://doi.org/10.1364/OPTICA.466337}{https://doi.org/10.1364/OPTICA.466337}}
\end{abstract}

\maketitle

\noindent Single, levitated mesoscopic objects are providing an excellent platform for sensing weak forces \cite{Ranjit2016} and exploring the classical-quantum boundary with massive objects \cite{Delic2020} due to their weak coupling to the environment and the ability to cool their centre of mass (CoM) motion. Intriguing possibilities have been proposed for interacting systems comprising multiple levitated particles, including quantum gravity measurements \cite{Marletto17}, dark matter detection \cite{Moore2021searching} and quantum friction measurement \cite{zhao2012rotational}, although there are currently few experimental demonstrations trapping multiple particles \cite{arita2018optical,svak2021stochastic,penny2021sympathetic,rieser2022tunable}.

The path to systems of interacting, massive quantum objects requires simultaneous cooling of multiple trapped particles. We take inspiration from experiments with cold neutral atoms \cite{Myatt97} and atomic ions \cite{larson1986sympathetic}, where an actively cooled object can be used to sympathetically cool another. Recently, sympathetic cooling of two charged particles held in a Paul trap was shown \cite{penny2021sympathetic}. While Coulomb forces provide strong interactions between particles, they also couple strongly to the environment. Hence, an all-optical alternative is desirable, especially for compatibility with state-of-the-art ground state cooling methods \cite{Delic2020}. 

\begin{figure}[htb!]
\centering
\includegraphics[width=1\columnwidth]{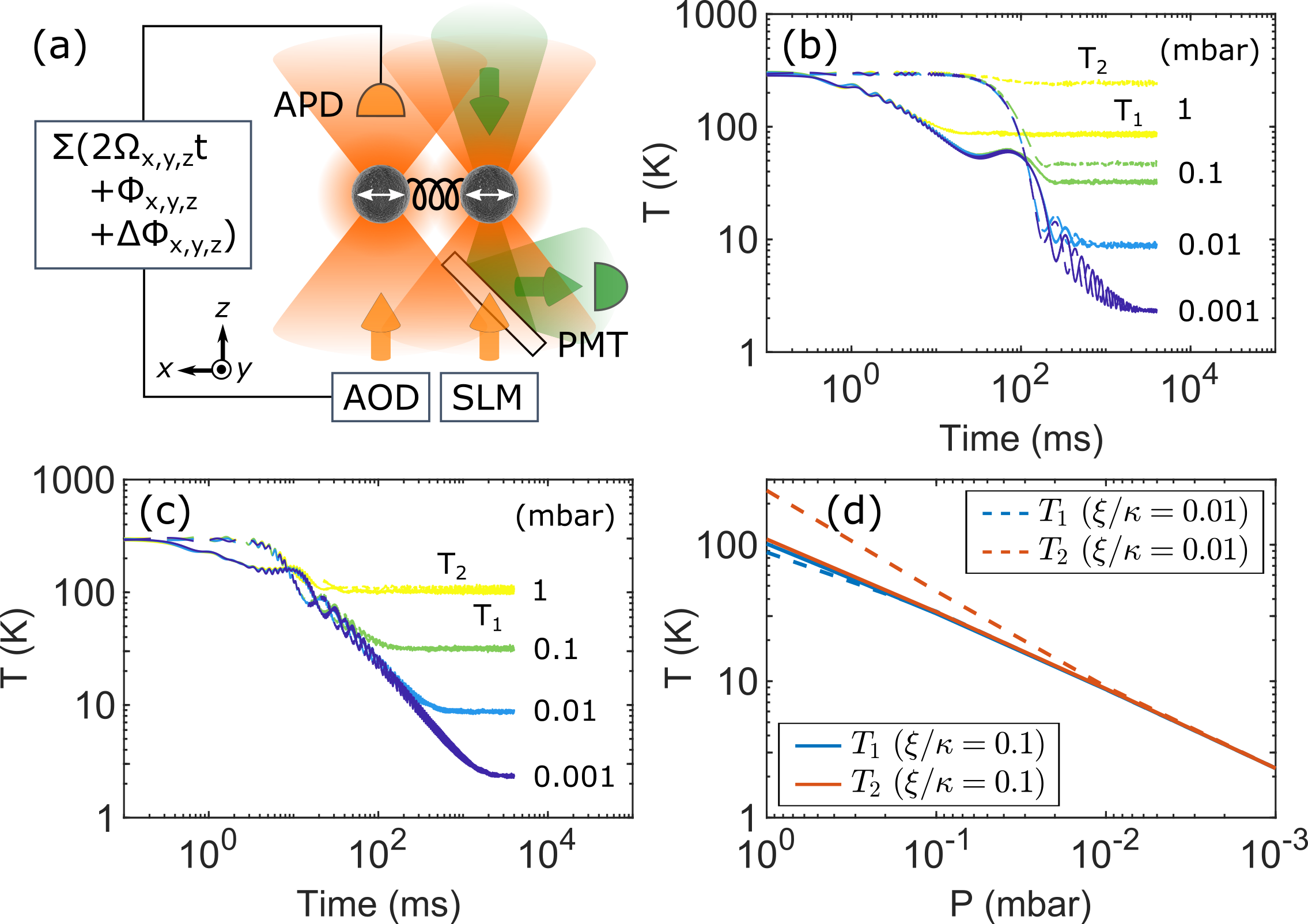}
\caption{
Sympathetic cooling scheme and numerical simulations. (a) Optical binding couples the centre-of-mass motion of two microspheres (depicted as a spring between the particles). When feedback cooling is applied to the left particle, the right particle is sympathetically cooled. (b-c) Time evolution of simulated centre-of-mass temperatures $T_1$ (solid lines) and $T_2$ (dashed lines) for the feedback-cooled and sympathetically cooled particle respectively, versus gas pressure for (b) $\xi/\kappa=0.01$ and (c) $\xi/\kappa=0.1$. (d) Simulated steady-state temperatures $T_1$ (blue) and $T_2$ (red) as a function of gas pressure for different binding strengths.}
\label{fig:schematic_sim_}
\end{figure}

Towards this goal, we have developed an optical tweezers system (Fig.~\ref{fig:schematic_sim_}(a)) that can trap more than one particle, mediate inter-particle separation and perform parametric feedback (PFB) cooling \cite{gieseler2012subkelvin} on one particle. Two rotating microparticles confined in separate but close-proximity traps exhibit optical binding, a light scattering mediated interaction whose strength is dependent on the inter-particle separation~\cite{arita2018optical}. We show, for the first time, the use of this optical binding to perform sympathetic cooling: by applying PFB cooling to one particle, the adjacent particle is sympathetically cooled to sub-Kelvin temperatures. 

We modelled the motion of two particles, optically bound along the x-axis and labelled $j=1,2$, using a Langevin equation for each particle \cite{arita2018optical}. The deterministic forces acting on the particles along the $x$-axis are $F_j=-\kappa x_j+\xi x_{3-j}$. Here, $\kappa$ is the trap stiffness for each individual particle, and $\xi$ describes the inter-particle coupling due to optical binding.  The ratio $\xi/\kappa$ (typically $\gtrsim0.1$ for vaterite microspheres) characterises the optical binding strength. To the above system of Langevin equations we have added PFB cooling applied to particle $j=1$ using the theoretical approach of Ref.~\cite{vovrosh2017parametric}.  PFB cooling in this case involves time-dependent modulation of the trap stiffness experienced by particle $j=1$, and is characterised by the ratio $\eta=\Delta\kappa/\kappa$ (typically $\lesssim0.05$), $\Delta\kappa$ being the modulation depth. 

Figures~\ref{fig:schematic_sim_}(b,c) show the time evolution of the simulated CoM temperatures for different binding strengths. PFB cooling decreases $T_1$ (solid lines), with the time to reach steady-state and the final temperature both depending on the pressure, due to the competition between PFB cooling and reheating from collisions with gas molecules. We quantify the sympathetic cooling effect using a measure of the temperature defined here as the ratio of the mean square displacement relative to that at room temperature. Temperatures below $1\,\mathrm{K}$ are attainable at $0.001\,\mathrm{mbar}$ gas pressure.  Similarly, the steady-state temperature of the sympathetically cooled particle, $T_2$ (dashed lines) depends on competition between sympathetic cooling and gas reheating. While the steady-state temperature is lowered for both optical binding strengths shown, for weak binding (Figure~\ref{fig:schematic_sim_}(b)) with $\xi/\kappa=0.01$ the particles do not equilibrate and $T_2>T_1$ at higher gas pressure. In contrast, for stronger optical binding (Figure~\ref{fig:schematic_sim_}(c)) with $\xi/\kappa=0.1$, $T_2\approx T_1$ at all pressures below $1\,\mathrm{mbar}$. Both particles can be cooled to below $1\,\mathrm{K}$ at $0.001\,\mathrm{mbar}$ gas pressure. Figure~\ref{fig:schematic_sim_}(d) depicts the steady-state CoM temperatures $T_1$ (blue) and $T_2$ (red) for PFB cooled and sympathetically cooled particles respectively, for different binding strengths $\xi/\kappa$. 

In our experiment, a continuous-wave (CW) laser at $1070\,\mathrm{nm}$ is used to create two circularly polarized optical traps (formed using a Nikon 100$\times$, NA=1.25 objective, total optical power $25\,\mathrm{mW}$) inside a vacuum chamber. The intensity of one of these traps is controlled by an acousto-optical deflector (AOD, IntraAction DTD-274HD6M) to allow PFB cooling. The position of the second trap, which holds the sympathetically cooled particle, is controlled by a spatial light modulator (SLM, Hamamatsu LCOS X10468-03) to set the optical binding strength. Two birefringent vaterite microspheres ($2.2\,\upmu\mathrm{m}$ in radius) are trapped at $9.8\,\upmu\mathrm{m}$ separation, where $\xi/\kappa\gtrsim0.1$. An avalanche photodiode (APD, Thorlabs APD410C) measures the CoM motion of \emph{only} the particle in the AOD trap. The APD signal is processed by a lock-in-amplifier (Zurich Instruments HF2LI) to modulate the AOD trap intensity by $\lesssim\pm5\,\%$ to realise PFB cooling of the CoM motion of this particle \cite{gieseler2012subkelvin}. The sympathetically cooled particle's CoM motion is measured by a photomultiplier tube (PMT, Thorlabs PMM01) using a separate CW beam at $532\,\mathrm{nm}$. 

\begin{figure}[htb]
\centering
\includegraphics[width=1\columnwidth]{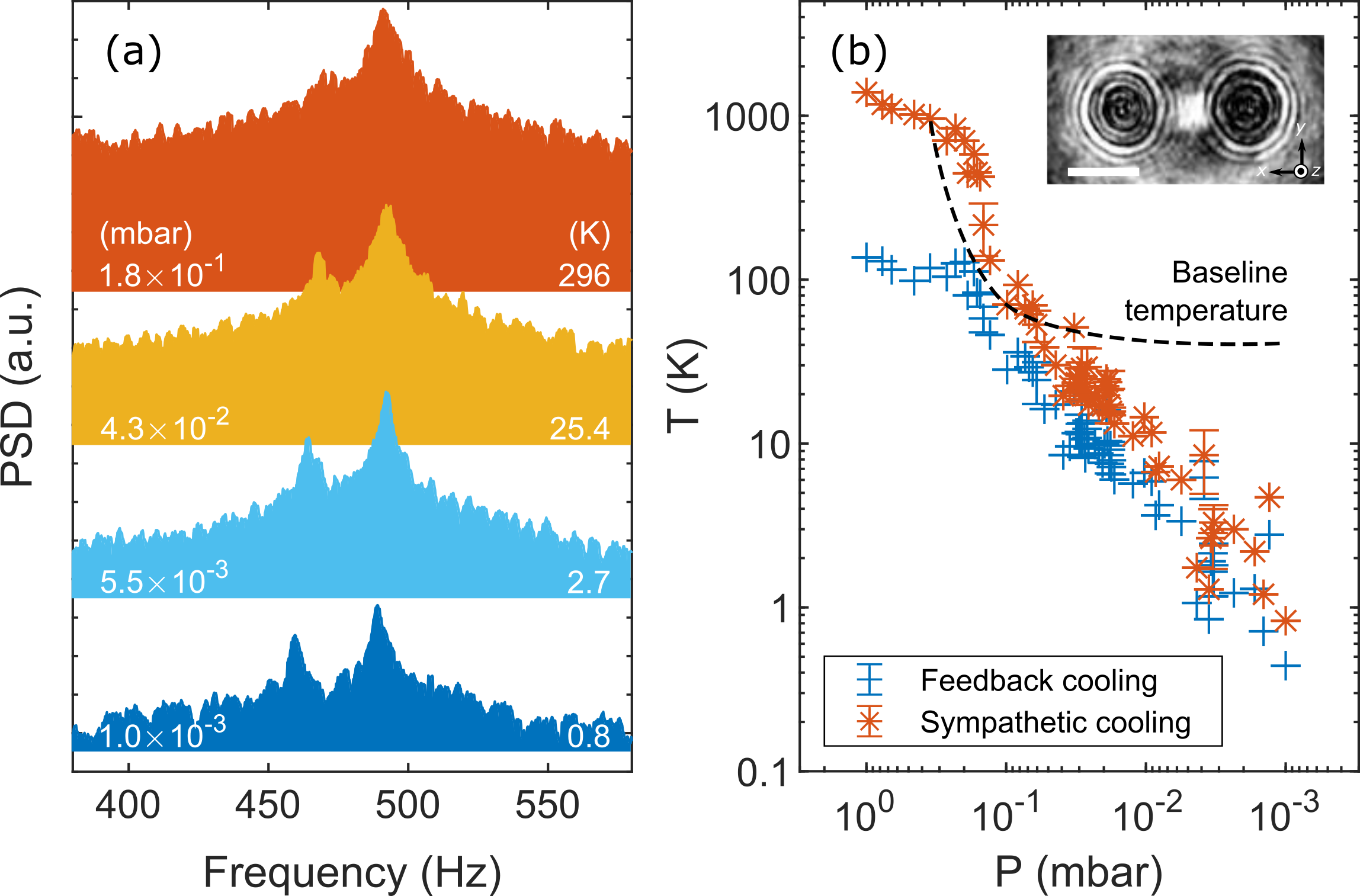}
\caption{
(a) Experimental observations of power spectral density of the sympathetically cooled particle at different gas pressures. These temperatures are measured while the PFB cooled particle is at $T=105\,\mathrm{K}$, $9.8\,\mathrm{K}$, $1.1\,\mathrm{K}$ and $0.4\,\mathrm{K}$, respectively, for each gas pressure. (b) Sympathetic cooling of a microsphere (red asterisks) via optical binding to the particle that is PFB cooled (blue crosses). The dashed line is a baseline curve for a single, rotating microsphere in the absence of PFB cooling. Inset: two trapped vaterite microspheres, where the scale bar shows $5\,\upmu\mathrm{m}$. 
}
\label{fig:two_particle_cooling_psds_offset_}
\end{figure}

Figure~\ref{fig:two_particle_cooling_psds_offset_}(a) shows the power spectral densities (PSD) associated with the CoM motion of the sympathetically cooled particle. The area under the PSD curve is proportional to the CoM temperature, which decreases as gas pressure lowers. Figure~\ref{fig:two_particle_cooling_psds_offset_} (b) shows the experimental results of the two particles' CoM temperatures depending on gas pressure: one cooled by PFB control (blue crosses) and the other sympathetically cooled via optical binding (red asterisks). We also show the baseline temperature for a single particle without PFB cooling (black dashed line). CoM temperatures are calibrated at room temperature with a gas pressure $>10\,\mathrm{mbar}$ without PFB. Both particles undergo significant cooling as the pressure is reduced below $\sim0.1\,\mathrm{mbar}$. In this regime, the rotation rate far exceeds the translational oscillation frequencies, and heating effects associated with resonant rotational-translational coupling are removed \cite{arita2013laser}. Both the PFB cooled particle and the sympathetically cooled particle can be seen to be cooled by two orders of magnitude compared to the baseline. While the lowest temperature of $\lesssim1\,\mathrm{K}$ agrees with the predictions in Figs.~\ref{fig:schematic_sim_}(b-d), non-conservative azimuthal forces can have significant effects on the dynamics of rotating levitated particles \cite{arita2022cooling}, and a more detailed model will be the subject of future work.

In summary, we demonstrate all-optical sympathetic cooling of optically trapped microspheres to sub-Kelvin motional temperatures.  In principle, future work with lower pressures and detection noise should allow the particles to be cooled even further. Such simultaneous cooling is a significant step towards the generation of cooled arrays of optically trapped particles, which will allow multi-particle studies at the quantum-classical boundary, including rotational quantum friction~\cite{zhao2012rotational}. 

\bigskip
The authors acknowledge funding from  Engineering and Physical Sciences Research Council (EP/P030017/1); Australian Research Council (DP220102303); Czech Science Agency (19-17765S); Ministerstvo \v{S}kolstv\'{i}, Ml\'{a}de\v{z}e a T\v{e}lov\'{y}chovy (CZ.02.1.01/0.0/0.0/15\_003/0000476).

The underlying data is available at \href{https://doi.org/10.17630/39f20d0d-74f1-4875-82b5-c2f61b038142}{https://doi.org/10.17630/39f20d0d-74f1-4875-82b5-c2f61b038142}.

\bibliography{main}
\end{document}